\documentclass{mn2e}
\usepackage{psfig}
\def\lesssim{{_ <\atop{^\sim}}}

\def\hkpc{h^{-1}\,{\rm kpc}}
\def\hMpc{h^{-1}\,{\rm Mpc}}
\def\hMsun{h^{-1}\,{\rm M_{\odot}}}

\def\fracj#1#2{{\textstyle{#1\over#2}}}
\def\b#1{{\bf#1}}
\def\d{{\rm d}}
\def\lesssim{\mathrel{\hbox{\rlap{\hbox{\lower4pt\hbox{$\sim$}}}\hbox{$<$}}}}
\def\gtrsim{\mathrel{\hbox{\rlap{\hbox{\lower4pt\hbox{$\sim$}}}\hbox{$>$}}}}

\begin{document}

   \title{Discreteness effects in cosmological N-body simulations}

   \author[J. Binney]
          {James Binney\\
           Theoretical Physics, 1 Keble Road, Oxford OX1 3NP\\
          }

   \date{Received ...; accepted ...}

   \maketitle

\begin{abstract}    
An estimate of the convergence radius of a simulated CDM halo is obtained
under the assumption that the peak phase-space density in the system is set
by discreteness effects that operate prior to relaxation. The predicted
convergence radii are approximately a  factor 2 larger than those estimated
for numerical convergence studies.

A toy model is used to study the formation of sheets of the cosmic web, from
which DM haloes form later. This model demonstrates the interplay between
phase mixing and violent relaxation that must also be characteristic of
spherical collapse. In the limit that sheets contain arbitrarily many
particles, it seems that power-law profiles are established in both distance and
energy. When only a finite number of particles is employed, relaxation is
prematurely terminated and the power laws are broken. In a given simulation,
the sheets with the highest peak phase-space densities are those that form
from the longest waves. Hence simulations with little small-scale power are
expected to form the cuspiest haloes.

\end{abstract}

\begin{keywords}
methods: $N$-body simulations -- galaxies: formation -- cosmology:
dark matter.
\end{keywords}

\section{Introduction}

Cold Dark Matter (CDM) is thought to dominate structure formation through
gravitational clustering. As its name implies, this material starts out very
cold, so that to an excellent approximation particles are confined to a
three-dimensional surface $[\b r,\b p(\b r)]$ in six-dimensional phase
space. The distribution function $f$ is infinite on this surface and zero
elsewhere.  The fine-grained distribution function is constant because it
obeys the collisionless Boltzmann equation. Hence, very high phase-space
densities are in principle possible right up to the present epoch. 

Studies of structure formation rely heavily on $N$-body simulations to
follow the clustering of dark matter (DM). Galaxies are believed to form
from baryons that have fallen in to the centres of DM haloes, so the
implications of various theories of DM for the observable properties of
galaxies depend sensitively on the small-scale structure of dark haloes.
Hence, it is important to understand how accurately $N$-body simulations
represent the inner parts of dark haloes, and this problem has attracted
considerable attention (Jing et al.\ 1995; Klypin et al.\
2001; Taylor \& Navarro 2001; Navarro 2003; Power et al.\ 2003; Fukushige,
Kawai \& Makino 2003; Hayashi et al.\ 2003).

On sufficiently small scales it is clear that the simulations will become
untrustworthy because they employ only a finite number of particles (Kuhlman
et al., 1996; Melott et al., 1997; Splinter et al., 1998; Binney \& Knebe, 2002).
Generally this scale is determined by resimulating a given halo with
improved resolution, in terms of the number of particles employed, the
number of waves used to define the initial conditions, the smallness of the
gravitational softening parameter $\epsilon$, and the smallness of the
timesteps. Based on such simulation series, Power et al.\ (2003) define a
radius $r_{\rm conv}$ outside which a given simulated halo should be
reliable, by calculating the smallest radius at which three criteria are all
met: (i) the local orbital time is much longer than the numerical timestep;
(ii) the centripetal acceleration of a circular orbit is smaller than
$0.5V_{200}^2/\epsilon$, where $V_{200}$ is the halo's characteristic
velocity; and (iii) the local two-body relaxation time is longer than the
Hubble time.

While it is clear that satisfaction of all three criteria is a necessary
conditions for a N-body simulation to be credible, it is not clear that it
is a sufficient condition: errors in the integration of the equations of
motion at any cosmic epoch can invalidate the final halo. We argue here that
a particularly important epoch to get right is the one that ends when
overdensities first collapse to form sheets that are virialized in only one
direction. We present evidence that discreteness effects at this primary
stage, rather than, for example, two-body relaxation in virialized haloes,
are what limit the central densities of simulated haloes.

We use a toy model and analytic calculations to clarify what happens when
virialized CDM structures form, and to understand the extent to which N-body
simulations of structure formation are compromised by artifacts due to
discreteness. The model explains why cuspy profiles form even when power at
high spatial frequencies is removed from the initial fluctuation spectrum.
It also suggests that in the absence of discreteness effects, virialized
haloes would have power-law cusps. Finally analytic arguments are used to
estimate the radius in a simulated halo within which the central profile is
flattened by discreteness effects that operate prior to virialization, and
we show that these estimates are in good agreement with the latest
numerical convergence studies.

\begin{figure*}
\centerline{\psfig{file=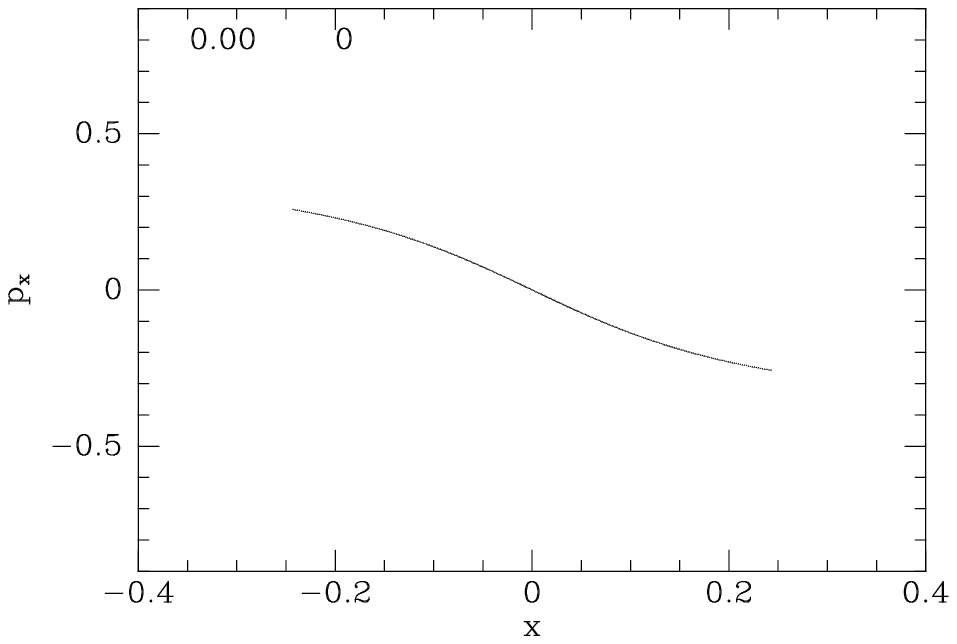,width=.46\hsize}\quad \psfig{file=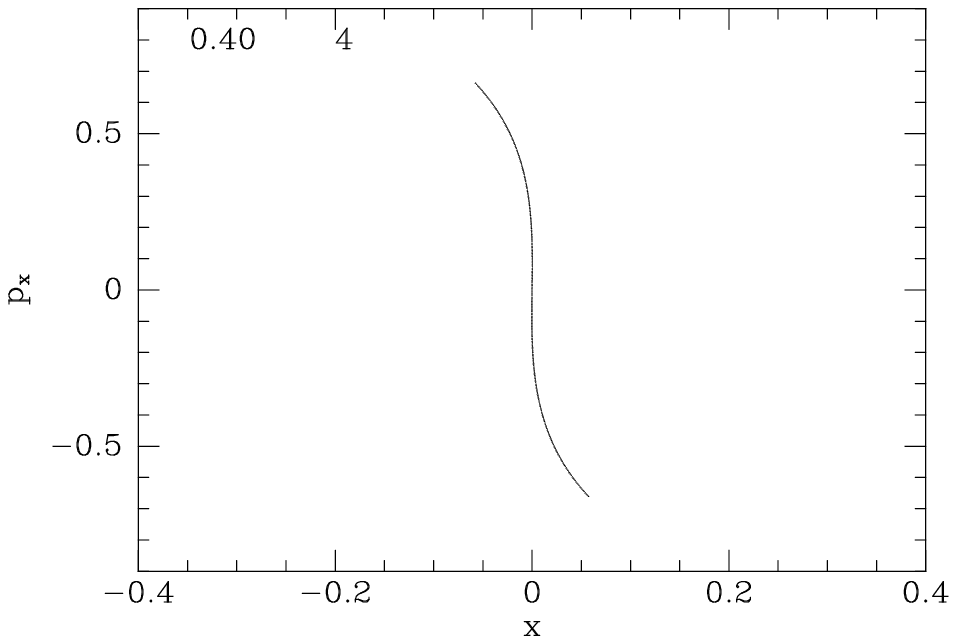,width=.46\hsize}}
\centerline{\psfig{file=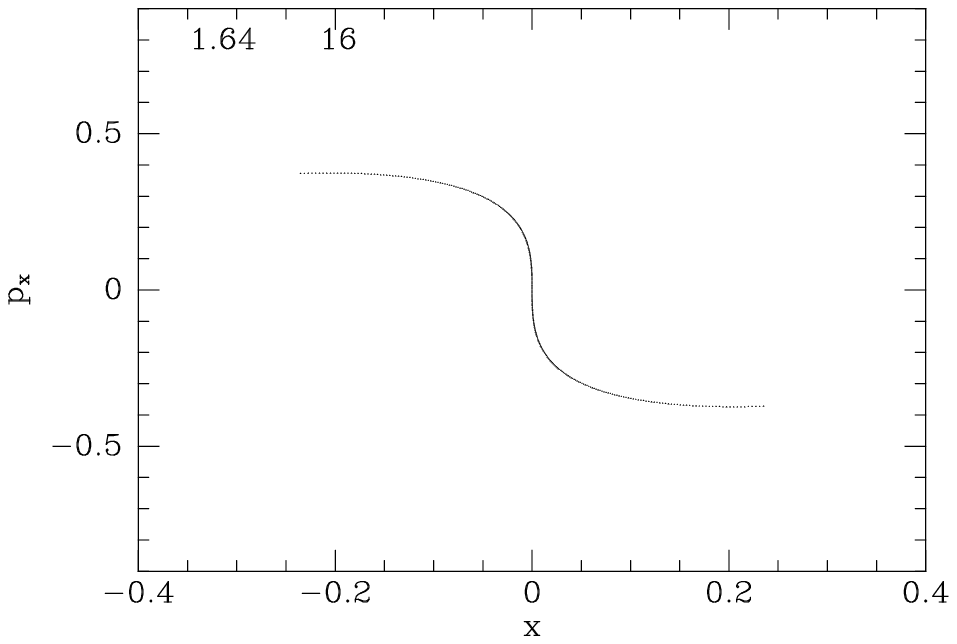,width=.46\hsize}\quad \psfig{file=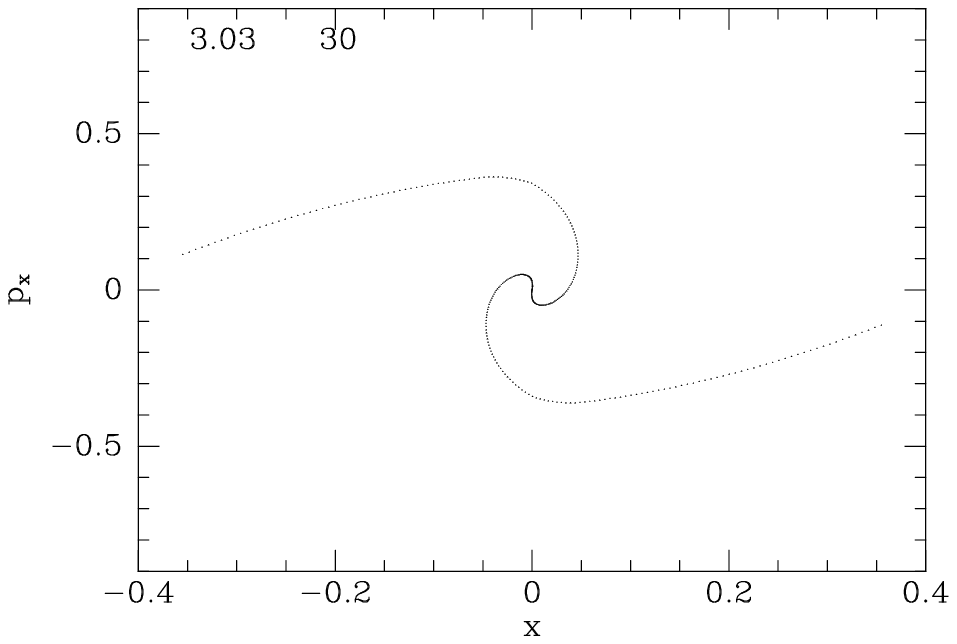,width=.46\hsize}}
\centerline{\psfig{file=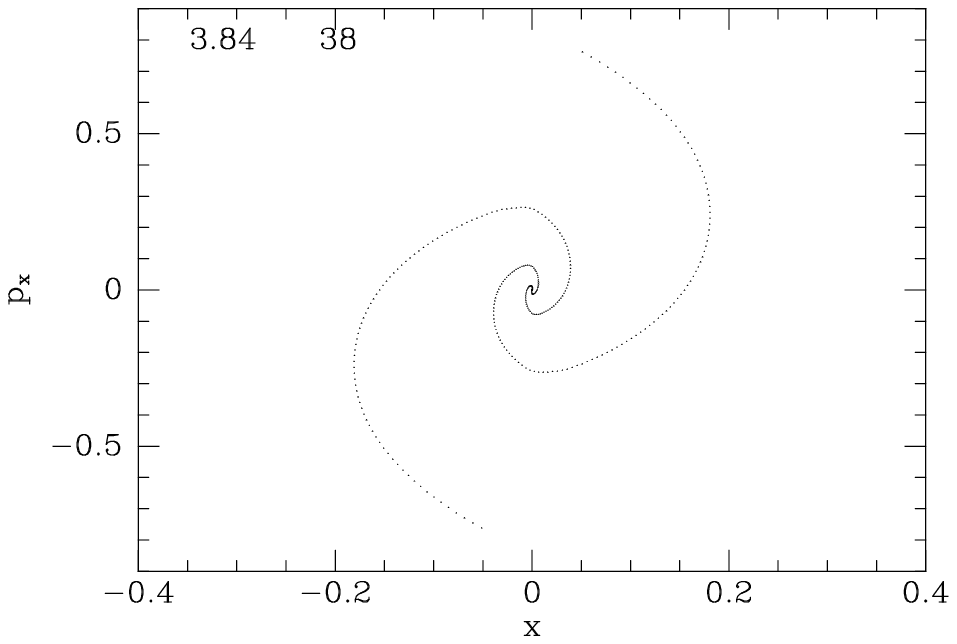,width=.46\hsize}\quad \psfig{file=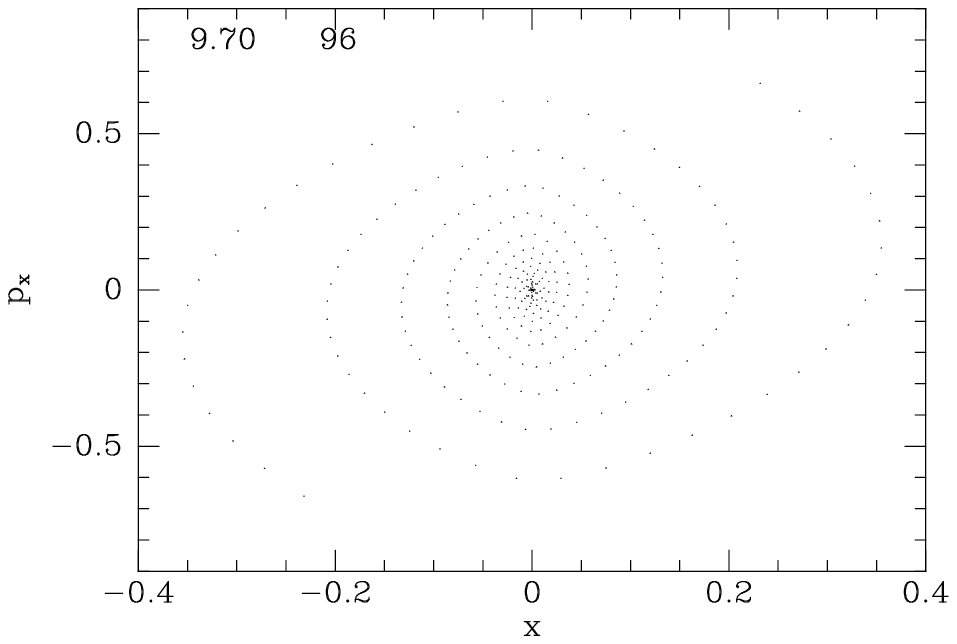,width=.46\hsize}}
\caption{From top left to bottom right, six stages in the violent relaxation
of 301 particles. The number in the top left corner of each panel is the time
since the initial conditions of equations (\ref{Xinit}) were imposed. The
units of time are the turnaround time $\tau$ of the underlying homogeneous
distribution of particles.\label{Nbdy1}}
\end{figure*}

\section{A toy model}\label{1dsec}

In this section we use a one-dimensional toy system to study the
interplay between phase mixing and violent relaxation, and to show how
discreteness cuts short this interplay. The simulations of Melott (1983) are
similar, except that they are  for hot dark matter and they employ a
conventional particle-mesh N-body technique to advance the particles.

Consider the dynamics of an odd number $2n+1$ of flat mass sheets that are
all perpendicular to the $x$ axis and move parallel to this axis. To avoid
confusion later, we henceforth call these sheets `particles'. The
particles, which each have surface density $\Sigma/n$, interact only through
gravity.  There is no net force on the middle particle. Let $k$ be the rank of
a particle with respect to this middle particle, with $k<0$ for particles on its
left.  Then the particle with rank $k$ experiences a force $F_k=- 4\pi
G\Sigma(k/n)$ that only changes when the particle passes
through another particle. Hence the particle's location, $x_k(t)$, is
quadratic in time between such passages. So we can analytically solve for
the instant at which the next passage occurs, and update the phase-space
coordinates of the particles to this time to machine precision. The ranks of
the passing particles are then exchanged and we solve for the instant of the
next passage. Hence this model provides a convenient laboratory for a
high-precision study of virialization.

We adopt initial conditions appropriate to the moment of turnaround of an
overdense region of the Universe. The ranking of the particles will not have
changed up to this time, so the requirement that at some fixed time $\tau$
in the past, all particles were located at the origin becomes
 \begin{equation}\label{chaindiff}
x_k-v_k\tau-2\pi G\Sigma(k/n)\tau^2=0\qquad(k=-n,\ldots,n).
\end{equation}
 This set of equations has a solution that corresponds to a homogeneous
distribution at turnaround
\begin{equation}
x_k=2\pi G\Sigma\tau^2(k/n),\quad v_k=0.
\end{equation}
 If we define $X_k=x_k-x_{k-1}-2\pi G\Sigma\tau^2/n$ and $V_k=v_k-v_{k-1}$,
 then  equations
(\ref{chaindiff}) yield a relation between $X_k$ and $V_k$, namely
 \begin{equation}
V_k={X_k\over\tau}.
\end{equation}
 Fig.\ref{Nbdy1} shows from top left to bottom right six stages in the
relaxation of 301 particles from the initial conditions obtained by choosing
 \begin{equation}\label{Xinit}
X_k=-0.6{\pi G\Sigma\tau^2\over n}\cos(0.3\pi k/n).
\end{equation}
 The time, in units of $\tau$, since these initial conditions were imposed
is given in the top left corner of each panel. The negative slope in the top
left panel implies that at the start of the simulation, the system is
already collapsing. In the next panel ($t=0.4$) the vertical orientation of
the distribution near the origin indicates that the centre has at that time
finished collapsing.  By the time, $t=1.64$, of the next panel, the centre
has expanded and collapsed once more. The edge, by contrast, is still
in full collapse. The following panels show that the phase lag $\psi$
between the oscillations of the centre and the outside grows rapidly. By the
time, $t=9.7$, of the last panel, $\psi$ is many $\pi$, and near the
centre it has become hard to follow the spiral of phase points.

\begin{figure}
\psfig{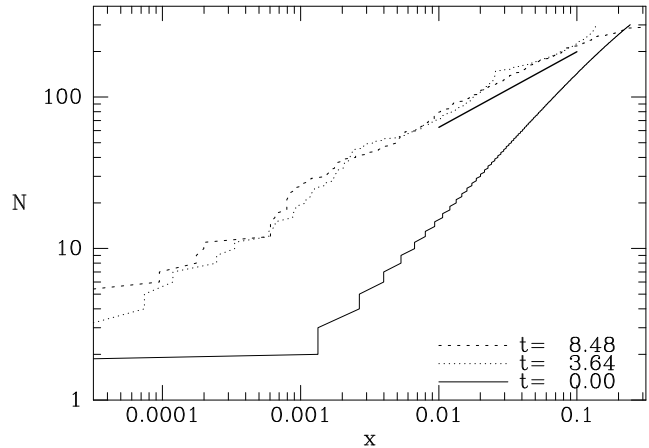}
\caption{The number of particles with $|x|$ less than a given value at three
times. The bold line shows the slope for
$N\propto|x|^{1/2}$.\label{plotdfig}} 
\end{figure}

Fig.~\ref{plotdfig} explores the evolution of the density profile of this
system by plotting the cumulative number of particles with $|x|$ less than any
given number at three times. In this logarithmic plot, the curve for $t=0$
has unit slope to within the errors because the density of particles
$\rho\sim\hbox{constant}$ initially. The slopes of the other two curves are
near ${1\over2}$, which implies $\rho\sim x^{-1/2}$. 

By Jeans' theorem, the distribution function of the equilibriated system must
be a function $f(E)$ of energy $E={1\over2}v^2+\Phi$.  From the fact that
the force law is 
 \begin{equation}\label{forcelaw}
F(x)=-4\pi G\Sigma(x/x_{\rm max})^{1/2},
\end{equation}
 where $x_{\rm max}$ is of order the amplitude of oscillation of the least
bound particles, it is easy to show that the power-law nature of the density
profile implies that
 \begin{equation}
f=f_0E^{-5/6},
\end{equation}
 where
 \begin{equation}
f_0=2^{-3/2}\left({\pi G\Sigma^4\over3 x_{\rm max}^2}\right)^{1/3}\left/
\int_0^\infty{\d y\over(y^2+1)^{5/6}}\right.\ .
\end{equation}

Two processes are manifest in Fig.~\ref{Nbdy1}. (i) The winding up of the
line of phase points into an ever tighter spiral is phase mixing. (ii)
Violent relaxation in the sense of Lynden-Bell (1967) causes the edge of the
occupied part of the phase plane to move outwards, and the points near the
centre to move towards the origin: the system's time-varying gravitational
field is transferring energy from the central to the peripheral particles.
The process by which this energy transfer occurs is this. Between the top
left and top right panels, the particles that are at $|p_x|\la0.2$ in the
second figure have fallen together under their mutual self gravity.
Particles further out have not been affected by this infall. However, when
the inner particles expand, the outer particles are falling in past them,
and the inner particles have to climb out of a deeper well than they fell into.
Conversely, the outer particles fall into a well that has significantly
weakened by the time they rise up the other side.

\begin{figure}
\psfig{file=plote.ps,width=\hsize}
\caption{Energy of the innermost 8 particles (full curve) and outermost
8 particles (dotted curve) as a function of time. For clarity the energy of
the inner particles has been multiplied by 2000.\label{plotefig}}
\end{figure}

\begin{figure*}
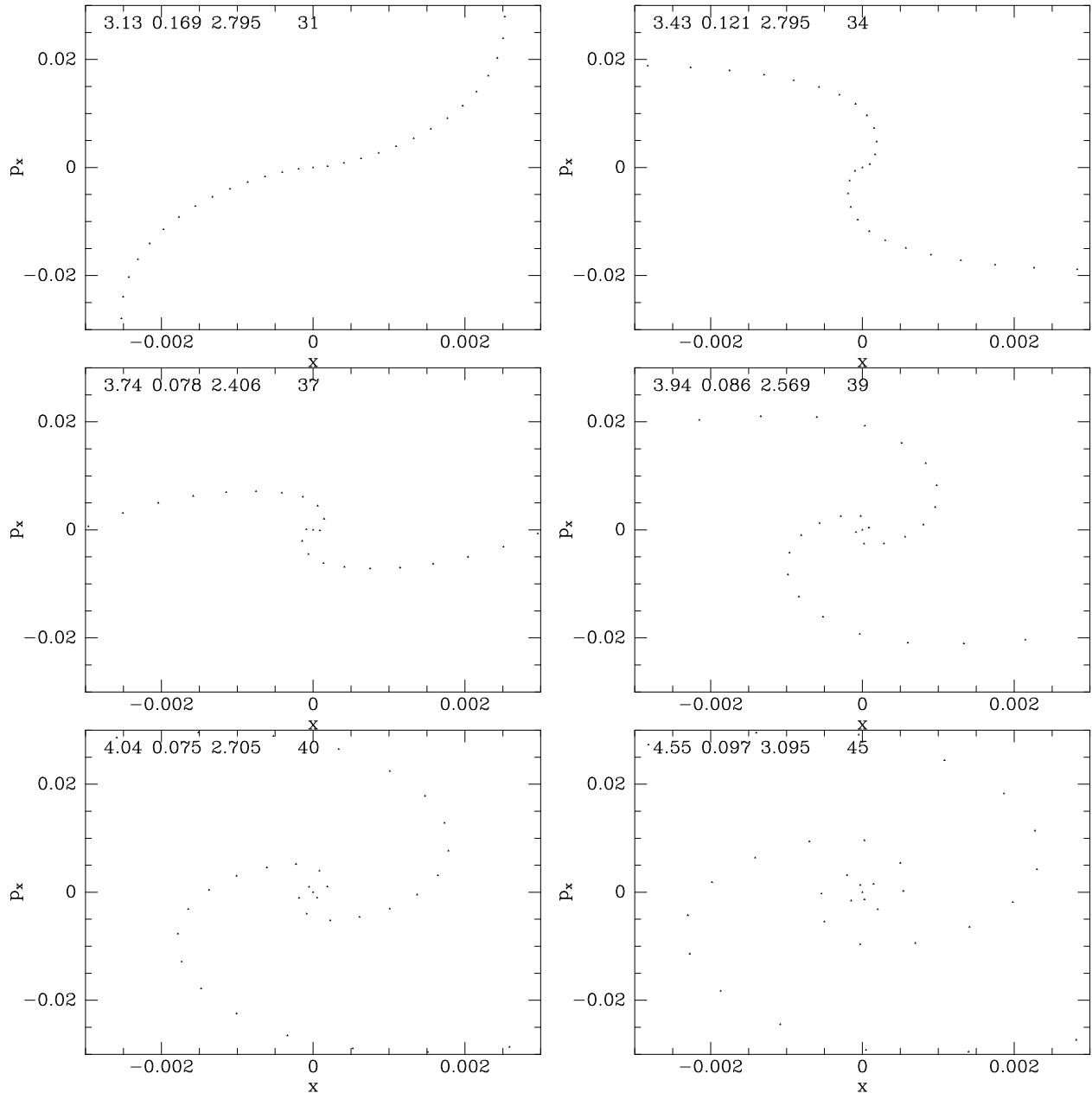

\centerline{\psfig{file=mid31.ps,width=.46\hsize}\quad \psfig{file=mid34.ps,width=.46\hsize}}
\centerline{\psfig{file=mid37.ps,width=.46\hsize}\quad \psfig{file=mid39.ps,width=.46\hsize}}
\centerline{\psfig{file=mid40.ps,width=.46\hsize}\quad \psfig{file=mid45.ps,width=.46\hsize}}
\caption{As Fig.~\ref{Nbdy1} but zoomed in on the centre of phase space and
at comparatively late times.\label{Nbdy2}}
\end{figure*}

This transfer of energy from the inner to the outer particles increases the
density contrast between the centre and the outside. Since the frequency of
a particle's oscillations through the centre scales as the square root of
the mean density $\overline{\rho}$ interior to it, the energy transfer
enhances the rate at which the function $\psi(E)$ is steepening. The energy
transfer works most effectively between groups of particles that differ by
$\sim\pi/4$ in $\psi$. So, as the first five panels of Fig.~\ref{Nbdy1}
clearly show, the characteristic distance scale of the transfer
rapidly decreases. In a useful, if oversimplified\footnote{The actual
numbers of particles (out of 301) losing energy at successive central
collapses are $\sim100$, $44$, $28$ and $19$.} picture, the first collapse
transfers energy between the inner and the outer half of the particles. The
second collapse at the centre transfers energy from the first quartile to
the second quartile, and a little later energy is transferred from the
second to the third quartile, and later still on out to the fourth quartile.
By this time the additional central collapse pictured in the fourth panel of
Fig.~\ref{Nbdy1} is transferring energy from the innermost ${1\over8}$ of
the particles, and so on. At each stage the transfers become smaller because
they involve a smaller and smaller fraction of the mass, but
Fig.~\ref{plotefig} shows that they remain significant to remarkably late
times.

Fig.~\ref{Nbdy2} shows  zoomed-in views of the centre of phase space at six
fairly late times. In the first three panels  the spiral of phase points is
unmistakable, but in the last three panels it has become hard to trace
because it bends through a large angle between points.
In these circumstances violent relaxation can no longer transfer energy
between particles, and the increase in the central mass density ceases.
Fig.~\ref{plotefig} confirms this conclusion by showing that the energy of
the innermost particles stops decreasing at about the time, $t=4.04$, of the
bottom left panel of Fig.~\ref{Nbdy2}.

The last panel of Fig.~\ref{Nbdy2} gives the impression that the system has
a finite central phase-space density. If we re-ran the simulation with more
particles, we would be able to trace the spiral even at this time, and the
impression of a finite central phase-space density would be dispelled.
Violent relaxation would continue to later times and generate a higher
central mass density. Fig.~\ref{plotefig} shows that once the phase-space
spiral has ceased to be resolved, the energy of the innermost particles
increases, presumably as a manifestation of two-body relaxation (i.e.,
stochastic heating).

In is interesting to quantify the peak phase space density. This will be of
order $\Gamma^{-1}$, where $\Gamma$ is the Poincar\'e invariant of the orbit
of particles with rank $|k|\sim2$.  From the force law (\ref{forcelaw})
the energy equation is
 \begin{equation}
{\textstyle {1\over2}}\dot x^2=4\pi G\Sigma{2\over3\surd x_{\rm max}}
\left(X^{3/2}-x^{3/2}\right),
\end{equation}
 where $X$ is the amplitude of a particle's oscillations. Equating $\Gamma$ to
the product of $X$ and the corresponding peak speed for orbits with $|k|=2$,
we find
 \begin{equation}\label{fmaxofn}
f_{\rm max}\sim{1\over\Gamma}
=\left({\textstyle{16\over3}}\pi G\Sigma x_{\rm max}^3\right)^{-1/2}\left({n\over2}\right)^{7/2}.
\end{equation}

This equation can be exploited in two ways. First we can consider the effect
of increasing the numerical resolution at which a given structure is
simulated. Then $n$ increases while the other factors are constant, so the
peak value of the DF rises steeply with $n$.

Alternatively, we can apply equation (\ref{fmaxofn}) to a series of
structures within a given simulation. Then $\Sigma$ and $x_{\rm max}$ are
both proportional to $n$, so overall $f_{\rm max}\sim n^{3/2}$. Hence the
highest phase-space densities are to be found in the largest structures,
because it is in these structures that violent relaxation works most
efficiently. 

\subsection{Case of finite initial phase-space density}

Consider the case in which the initially populated region of phase space has
a finite width -- as in the case of warm dark matter. The winding up of the
spiral and violent relaxation will proceed as before until the spiral's
radius of curvature becomes comparable with the width of the populated
sheet. At this point violent relaxation ceases to be possible. Hence, the
peak phase-space density of the relaxed system will be comparable to the
initial phase-space density of the sheet. Melott (1983) reached a similar
conclusion on the basis of his simulations of pancake formation in a
hot-dark-matter cosmology.

\subsection{Summary of experimental results}

From our one-dimensional example we conclude the following

\begin{enumerate}
\item Material at each distance from the centre executes oscillations. The
phase of these oscillations becomes a progressively steeper function
$\psi(E)$ of distance from the centre of phase space.

\item Gravity transfers energy with great efficiency between particles whose
oscillations are out of phase by $\sim\pi/4$. Even a small initial
enhancement of density towards the centre causes this transfer to be outward.

\item The outward transfer of energy enhances the density contrast, which
steepens the function $\psi(E)$, and reduces the length scale over
which individual energy transfers occur.

\item Outward energy transfer continues until the underlying spiral in phase
space ceases to be resolved in the sense that the spiral turns through a
large angle between phase points.

\item In the limit of an arbitrarily large number of particles, the spiral remains resolved
through arbitrarily many windings, and the density profile of the final
relaxed structure is strictly singular.

\item The large dynamic range between the smallest lengthscale of energy
transfer and the overall system size, combined with the hierarchical nature
of the energy transfer process serve to establish a power-law density
profile $\rho\sim x^{-1/2}$ and a coarse-grained phase-space density $f\sim
E^{-5/6}$.

\item Once the phase-space spiral has ceased to be resolved, or its width is
no longer negligible, a core is established and two-body relaxation causes
the halo to heat the core.

\end{enumerate}

\section{Relevance for cosmology}

Has the toy model any relevance for realistic three-dimensional systems? The
interplay between phase mixing and violent relaxation that it so clearly
illustrates is certainly generic and as such relevant to nearly spherical
systems.  The period at which a star oscillates radially in a spherical
potential is strictly a function of both energy and angular momentum.
However, in a typical galactic potential the radial period of a star is
largely determined by the star's energy.\footnote{The dependence on $L$
vanishes entirely in the case of the isochrone potential.} Consequently,
motion in the $(r,p_r)$ phase plane is closely analogous to the motion in
the $(x,p_x)$ plane studied in the last section.  The shells or ripples that
are seen in unsharp-masked images of a significant fraction of elliptical
galaxies (Malin \& Carter 1983) are  manifestations of the
winding up of an initially narrow distribution of stars in the $(r,p_r)$
phase plane (Hernquist \& Quinn 1988) very much like what we see in
Fig.~\ref{Nbdy1}. 

As Zel'dovich (1970) pointed out, the non-linear stage in cosmic structure
formation is initially a one-dimensional process. It gives rise to the
network of very overdense sheets that are now familiar as the `filaments' of
the `cosmic web' that are seen in slices cut through computer simulations of
gravitational clustering (Shandarin et al., 1995). The formation of such a
virgin sheet will closely parallel the dynamics of the last section.
Differences of detail will arise early on through the continued transverse
expansion of the matter, which makes the force acting between two wafer-thin
slices of matter an explicit function of time, but these differences are
likely to be unimportant, given the finite duration of the one-dimensional
relaxation process, and the tendency of the motion in the perpendicular
directions to be faltering and turning to contraction soon after a
singularity is reached in the first direction. 

Thus virgin sheets of the cosmic web will have singular central densities
$\rho\sim x^{-1/2}$. The characteristic transverse length scale of these
first sheets depends upon the power spectrum of the initial fluctuations.
Unless the power spectrum has a sharp cutoff, many different length scales
will characterize the transverse structure of the virgin sheets within a
sufficiently large volume of space because many different linear modes
combine to determine how and when a given part of the Universe will go
non-linear and collapse.

Eventually the sheets collapse transversely, either upon themselves or
through falling onto the massive objects that tend to form at the nodes of
the original cosmic web. During this second stage of structure formation,
violent relaxation can occur again. It may be possible to idealize the
dynamics as occurring in the phase plane defined by the direction of fastest
infall, $y$, and its conjugate momentum. In this plane an initially narrow
distribution of phase points will be wound up into a spiral, and the final
density profile is likely to be a cusped function $\rho_y\sim y^{-\beta}$
with $\beta>0$. The strong initial inhomogeneity of the system will result
in the spiral winding up much more rapidly than it does when a virgin sheet
forms, so we expect $\beta<{1\over2}$. Finally the over-dense tube resulting
from collapse in the $x$ and $y$ directions will collapse along its length,
$z$. Again a cusped final density profile in $z$ is to be expected,
$\rho_z\sim z^{-\gamma}$, but violent relaxation will not be efficient, so
$\gamma<{1\over2}$. The final density profile $\rho(r)$ can be thought of as
the spherically averaged product of the profiles established in each
collapse direction.  It is easy to show that $\rho\sim r^{-(\alpha+\beta+\gamma)}$, where
$\alpha={1\over2}$. The exponent of $r$ must thus be greater than ${1\over2}$ and
is expected to be significantly less that ${3\over2}$. Simulations of cosmic
structure formation favour exponents that lie towards the upper half of this
expected range (Hayashi et al.\ 2003).

\subsection{Discreteness in cosmic N-body simulations}

We have argued that when a sheet of zero thickness in phase space collapses,
the final virialized structure should have a singular central density. We
have also shown that there is a maximum phase-space density that can be
achieved when a finite number of particles is used to sample the phase-space
sheet [eq.~ (\ref{fmaxofn})]. Here we ask what the maximum phase-space
density is that can be achieved in the best current N-body simulations of
cosmic structure formation, and inside what radius within a DM halo this
restriction becomes important.

We have seen that one of two effects can set an upper bound on the
phase-space density that occurs in a simulation. One is due to the
termination of violent relaxation by discreteness. For the one-dimensional
case it is quantified by eq.~(\ref{fmaxofn}). The other is set by any upper
limit on the initial phase-space density of the initial conditions, since
Liouville's theorem forbids any increase in the phase-space density.

It is not evident how to extend eq.~(\ref{fmaxofn}) to the three-dimensional
case. This is unfortunate, for its implication that the highest phase-space
densities are associated with the most massive objects, is unexpected. It
also needs clarification: it does not apply to objects that form through the
merger of virialized pieces. Hence, when there is considerable small-scale
power, few if any objects to which it applies will have survived to the
present epoch. However, the result does explain why Knebe et al.\ (2002)
found that suppression of small-scale power in the input spectrum did not
reduce the cuspiness of the final haloes. Also, it suggests that the material
that now forms the cores of simulated haloes virialized in narrow ranges of
length-scale and redshift: this material was in the largest structures to
collapse before substructure could form within them.

In view of the difficulty of generalizing equation (\ref{fmaxofn}) to three
dimension, we concentrate on ways in which discreteness effects limit the
phase-space density achievable prior to virialization. Below we show that
the upper limit on $f$ that we derive from these effects predicts
convergence radii for DM haloes in simulations that are in reasonable
agreement with those determined empirically. This finding suggests that
restrictions on $f$ prior to virialization are dominant in practice.

Prior to virialization discreteness limits $f$ in two ways:
\begin{enumerate}
\item With a small number of sampling points, it is not possible accurately
to trace a curved phase-space sheet.

\item In simulations the gravitational forces are estimated by concentrating
mass into lumps. This concentration adds high-frequency noise to the force
field. Neighbouring particles sample this noise in largely uncorrelated
ways, and thus acquire uncorrelated components of momentum that are 
discreteness artifacts. These momentum components broaden an initially
perfectly thin sheet of phase points.

\end{enumerate}

We estimate these effects in  Appendix A. The second effect is estimated in
two complementary ways. In every case we estimate the Poincar\'e
invariant $\Gamma$ over which two or three particles are distributed in one
pair of canonically conjugate coordinates. Since there are three such pairs,
our estimate of
the DF is
 \begin{equation}
f\sim{m\over\Gamma^3},
\end{equation}
 where $m$ is the mass of a simulation particle. 

Equations (\ref{firstest}), (\ref{unequiest}), and (\ref{grainest}) give
three estimates of $\Gamma$.  The expressions all contain the dimensional
factor $H_0(\Delta q)^2$, where $H_0$ is the Hubble constant and $\Delta q$
is the comoving separation of particles at early times. This factor is
multiplied by a numerical factor that is comparable to unity, but probably a
little smaller. Thus below we use the estimate
 \begin{equation}\label{basicfest}
f\sim{\eta m\over[H_0(\Delta q)^2]^3},
\end{equation}
 where $\eta $ is a factor a little smaller than unity.  Equation
(\ref{givestau}) and equation (\ref{grainv}) independently imply that $f\sim
a^{-9/2}$ in the run up to virialization.

When the system virializes, we have argued that the peak phase-space density
will decrease only if the phase-space sheet is thinly sampled compared to
its width. In Appendix B we show that on virialization the peak value of the
DF does not diminish much below that given by equation (\ref{basicfest}).

\section{Application to simulations}

Hayashi et al.\ (2003) studied the convergence of the central structure of DM
haloes in state-of-the art simulations of structure formation. It is
interesting to compare the convergence radius $r_{\rm conv}$ that they
derived empirically from these simulations with the one that follows our
considerations.

Let us assume that DM haloes have cores in which $\rho(r)=\rho_0(
r/r_0)^{-1}$ and the velocity dispersion tensor is isotropic -- simulations
suggest that these assumptions are not seriously violated. Then, with the
gravitational potential taken to be zero at the centre, the distribution
function of the system is easily shown to be
 \begin{equation}
f(E)={3G\rho_0^2r_0^2\over(2E)^{5/2}}.
\end{equation}
 We find the radius $r_{\rm min}$ at which the DF from
this formula at $v=0$ equals the maximum phase density of
equation (\ref{basicfest}). We find
 \begin{equation}
r_{\rm min}^{5/2}={3[H_0(\Delta q)^2]^3\over
32\pi^{5/2}G^{3/2}(\rho_0r_0)^{1/2}\eta m}.
\end{equation}
 As one proceeds inwards from $r_{\rm min}$, more and more of the density is
contributed by parts of phase space at which the DF exceeds the value of
eq.~(\ref{basicfest}). In a simulation these portions of phase space are
under-populated, so the density falls more and more below its true power-law
value. Hence $r_{\rm min}$ is our prediction of the convergence radius of
a simulated halo.

 If $N^3$ particles are used to represent the matter in a box of side $L$,
the mass of an individual particle is
\begin{equation}
m=\Omega_{\rm m}{3H_0^2\over8\pi G}\left({L\over N}\right)^3,
\end{equation}
 and $r_{\rm min}$ satisfies
\begin{equation}
r_{\rm min}^{5/2}={H_0L^3\over
4\pi^{3/2}(G\rho_0r_0)^{1/2}N^3\eta\Omega_{\rm m}}.
\end{equation}
 We apply this formula to the G1 sequence of models listed in Table 1 of
Hayashi et al. For the relevant halo
$\rho_0r_0=6.34\times10^7\hMsun(\hkpc)^{-2}$, while $L=5\hMpc$, so
\begin{equation}
r_{\rm min}=2.14(\eta\Omega_{\rm m}/0.3)^{-2/5}(N/256)^{-6/5}\hkpc.
\end{equation}
 As one proceeds from $N=32$ to $N=256$ down the G1 sequence of models, we
find that $\eta^{2/5}r_{\rm min}/r_{\rm conv}=1.14$, $1.47$, $1.53$, and
$1.53$.  So $r_{\rm min}$ is larger than $r_{\rm conv}$ by a factor
$\sim1.5\eta^{-2/5}$ that is only slightly greater than unity. This
agreement between our analytic estimates and the numerical studies suggests
that the peak densities within simulated haloes are indeed set by
discreteness effects that operate before virialization.

\section{Conclusions}

Semi-analytic work on the formation of cosmic structure usually focuses on
the collapse of spherical over-densities (Gunn \& Gott 1972; Fillmore \&
Goldreich 1984; Bertschinger 1985; Subramanian, Cen \& Ostriker, 2000).
Actually one direction collapses before the other two, so the crucial first
phase of the virialization process is one-dimensional (Zel'dovich, 1970;
Shandarin et al., 1995).  Fillmore \& Goldreich (1984) obtained a
self-similar solution of the collapse problem in this planar symmetry.
Unfortunately, their results are not directly comparable to ours because (i)
they assumed that the transverse expansion continued unperturbed, and (ii)
their initial overdensity was constrained to diverge at the centre of the
sheet.  Consequently, for all $t>0$ their solution has a
collapsed core, and the phase-space spiral has a time-invariant form: its
scale changes with time, but it does not wind up in the manner of
Fig.~\ref{Nbdy1}.

Our toy model of the planar collapse of CDM reveals a strong interplay
between phase mixing and violent relaxation, in which an initially small
phase lag of the oscillations between the centre and periphery of an
overdensity causes energy to flow outwards, and this outward energy flow
hastens the growth of the phase difference, which in turn causes the
exchange of energy to repeat on ever smaller scales. The numerical evidence
suggests that in the limit of arbitrarily many particles, power-law profiles
in space and energy would be established as a natural consequence of the
large dynamic range between the smallest scale of energy interchange and the
overall size of the system. Power laws are also consistent with the original
infinite phase-space density persisting at the centre of the relaxed system,
while admixture of `air' ensures that the phase-space density is finite and
outwards-decreasing elsewhere.

When the system is represented by only a finite number of particles, random
motions develop prior to virialization. Consequently, as collapse begins,
the phase-space density is finite and the interplay between phase mixing and
violent relaxation is prematurely terminated. The central phase-space
density of the relaxed system is comparable to the phase-space density just
before collapse, while further out it differs little from the value that
would hold in the limit of an arbitrary number of particles. Consequently,
the real-space density profile must turn over near the centre from the slope
it would have in that limit.

An interesting result from the one-dimensional model is that, in a given
simulation, the peak phase-space density in simulated sheets of the cosmic
web scales with particle number as $f\sim n^{3/2}$. Hence, the sheets with
the highest surface density achieve the largest phase-space density. The
surface density of the sheets that form is increased if we suppress
small-scale power in the initial fluctuation spectrum, so there is a
suggestion that simulations that have little small-scale power will produce
the cuspiest haloes. This finding is consistent with the counter-intuitive
results of Knebe et al.\ (2002).

The extension to the formation of DM haloes of these results for the
formation of the sheets of the cosmic web is not straightforward. Some
insight can be gained by imagining that the original process repeats as each
of the other directions experiences collapse. This idealization is
problematic because the collapses of these directions are rarely well
separated in time. However, it is probably legitimate to argue that in the
limit of infinite particle number, the final density profile should be a
power law\footnote{Figure
4 of Hayashi et al.\ (2003) argues against this conclusion, however.}
because the collapsing sheet brings no relevant scale into the
relaxation process, and the latter has no intrinsic scale. Also
it seems likely that violent relaxation will be less efficient in later
collapses than it was in the first, because the phase spirals associated
with these directions will wind up very quickly. This reasoning allows us to
set an upper limit ${3\over2}$ on the power-law index of the final density
cusp, which is clearly greater than ${1\over2}$.

If we assume that in the limit of infinite particle number the slope would
be $\alpha=1$, as in the NFW profile, then we can estimate the radius
$r_{\rm min}$ at which the maximum phase-space density sampled equals the
phase-space density that developed prior to virialization. This radius forms
a reasonable estimate of the radius at which the simulated density profile
turns away from a line of unit slope. We find that $r_{\rm min}$ is larger
by a factor $\sim2$ than the convergence radius $r_{\rm conv}$ of Hayashi et
al.\ (2003). This finding strongly suggests that processes that take place
prior to virialization determine the smallest radius at which a simulated
halo can be trusted. 

The importance of knowing which structures in cosmological simulations are
reliable, is clearly very great. Brute-force convergence studies such as
those of Power et al.\ (2003) do not constitute a straightforward tool for
validating N-body simulations because the limit in which reality is
approached involves sending three quantities to zero: the mass of an
individual particle, the softening length, and the wavelength of the
highest-frequency mode included in the initial conditions. Before an
entirely convincing convergence study can be made, one needs to know along
which path in the three-dimensional space of these parameters one should
approach the origin (e.g., Splinter et al.\ 1998). Hitherto there has been a
tendendency for N-body simulators to focus on the impact that discreteness
has on the dynamics of relaxed systems. This study suggests that dicreteness
may have its dominant impact prior to virialization.

\section*{Aknowledgments}

This work was started in 2000 during a visit to the University of Washington
in Seattle and was stimulated by discussions with George Lake.  I am
grateful to Tom Quinn and Yago Ascasibar for comments on drafts of the paper.

\appendix
\section{Width of phase sheet}

The initial conditions for cosmological simulations are invariably set up by
superposing a number of plane-wave perturbations to the density and velocity
fields. Comoving coordinates are generally employed rather than the inertial
coordinates employed in  Section \ref{1dsec}.  In canonically conjugate comoving
coordinates, Zel'dovich's (1970) analytic solution for the evolution of an isolated
wave is
 \begin{eqnarray}\label{basice}
x&=&q+F\nonumber\\
p&=&a^2HF,
\end{eqnarray}
 where  $H=\dot a/a$ is the Hubble constant at scale factor $a$ and we adopt
\begin{equation}
F=-{a\over a_0k}\sin(kq).
\end{equation}
 Here $q$ is a Lagrangian coordinate and $a_0$ is the scale
factor at which the wave breaks and a caustic forms. 

\begin{figure}
\psfig{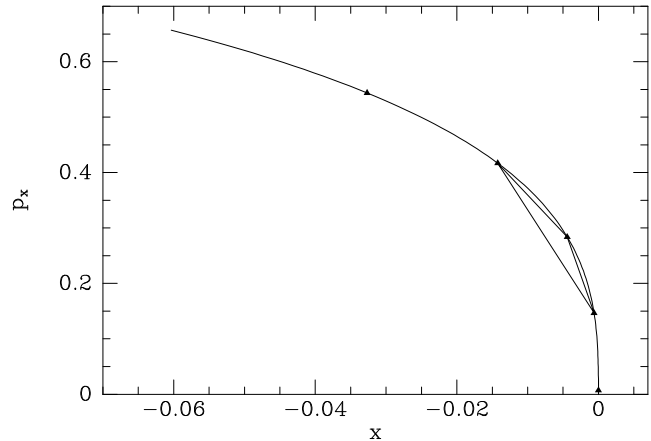}
\caption{An underlying Zel'dovich wave.\label{pointsfig}}
\end{figure}

\subsection{Effective width of a distorted sheet}

The phase-space density of a group of particles is  calculated by
evaluating the phase-space volume that they occupy. By definition, the
latter is the product of three Poincar\/e invariants $\Gamma_i=\int\d x_i\d
p_i$, one for each index $i$, with $p_i$ the momentum conjugate to $x_i$.
Consider therefore Fig.~\ref{pointsfig}, which shows the track in the
$(x_1,p_1)$ plane of a collapsing wave, and a number of
particles that are intended to realize this wave in a simulation. Any three
consecutive particles define a triangle. We now evaluate the area of this
triangle, which is the Poincar\'e invariant that the particles define. 

Taking the modulus of the vector product
$(x_1-x_2,p_1-p_2)\times(x_2-x_3,p_2-p_3)$ of two differences between three
vectors $(x,p)$ of the form (\ref{basice}), we find the Poincar\'e invariant
 \begin{equation}
\Gamma={Ha^2\over2}[(q_1-q_2)(F_2-F_3)-(q_2-q_3)(F_1-F_2)]
\end{equation}
 If we now expand $F_1$ and $F_3$ as Taylor series around $q_2$, we find to
 leading order
\begin{eqnarray}\label{givestau}
\Gamma&=&\fracj14Ha^2F''(q_2)(q_2-q_1)(q_3-q_2)(q_3-q_1)\nonumber\\
&=&{Ha^3k\over2 a_0}\sin(kq)(\Delta q)^3,
\end{eqnarray}
 where in the last line we have assumed that the particles are uniformly
spaced in $q$ with interparticle separation $\Delta q$.  In a flat universe,
$H=a^{-3/2}H_0$, where $H_0$ is the Hubble constant at the current epoch,
and we see from equation (\ref{givestau}) that $\Gamma\sim a^{3/2}$.  This
estimate of $\Gamma$ is credible up to the epoch of virialization, $a=a_0$.
Hence, at virialization the particles occupy a minimum Poincar\'e invariant
 \begin{equation}\label{firstest}
\Gamma=H_0(\Delta q)^2\,[\fracj12(k\Delta q)a_0^{1/2}\sin(kq)].
\end{equation}
 This expression for the Poincar\'e invariant is made up of a dimensional
factor
 \begin{equation}\label{dimless}
\Gamma_0=H_0(\Delta q)^2,
\end{equation}
 and a dimensionless factor
\begin{eqnarray}
\eta&=&\fracj12(k\Delta q)a_0^{1/2}\sin(kq)\nonumber\\
&\simeq&\fracj12(k\Delta q)^2a_0^{1/2}
\end{eqnarray}
 where the second line follows because a wave breaks, and a high-density
virialized structure forms, around $q=0$.  At the Nyquist frequency,
$k\Delta q=\pi$, so this factor will be of order, but less than unity is
cases of interest. The other two factors are each $\sim\fracj12$.  Hence,
equation (\ref{dimless}) gives a useful order-of-magnitude estimate of the
smallest achievable value of $\Gamma$.

\subsection{Unequilibriated fluctuations}

The process of setting up a cosmological simulation can be conceptualized as
a two-stage process. One first sets up a discrete representation of a
perfectly homogeneous distribution of matter, and then one perturbs it by
displacing particles in a systematic way that generally involves waves of
some kind. Let us focus on potential fluctuations implicit in the
unperturbed `homogeneous' distribution of particles, since these
fluctuations are clearly an artifact due to discreteness.  The nature of
these fluctuations is elucidated by the `glass' approach to realizing a
homogeneous matter distribution. In this technique (White, 1996) particles
are first distributed randomly within the simulation volume and are then
moved subject to the usual equations of motion with gravity made into a
repulsive force. After a sufficient number of integration steps, each
particle comes to rest at the bottom of its nearest potential well. When
gravity is restored to an attractive force, the potential well becomes a
hill with the particle at its summit. When particles are distributed on a
lattice, they again arrive at the summits of hills, and the structure of
these hills can be investigated by Ewald summation.

When the `homogeneous' configuration is perturbed, the hills are distorted,
but only mildly if, as should be the case, the perturbation is imposed at
sufficiently large redshift for the imposed displacements to be small
compared to the inter-particle separation. Since the structure of an
individual hill depends on the positions of {\em all\/} particles, each
particle is displaced slightly from the summit of its own hill, and will
roll down the hill as soon as the simulation is integrated forward in time.
Its rolling motion is nothing but a discreteness artifact and will cause the
Poincar\'e invariant associated with neighbouring particles to grow as
$p_{\rm r}\Delta q$, where $p_{\rm r}$ is the momentum of a typical
particle's roll.

Let the parabolic form of the potential around the summit be 
 \begin{equation}\label{defsPhi0}
\Phi(x)=-{\Phi_0\over2(\Delta q)^2}x^2,
\end{equation}
 where $\Phi_0$ is an appropriate constant. Then from the Hamiltonian
$\fracj12 p^2/a^2+\Phi/a$ we easily find that $x$ satisfies the equation of
motion  
\begin{equation}
x''+\fracj32{x'\over a}-{\Phi_0\over(H_0\Delta q)^2}{x\over a^2}=0,
\end{equation}
 where a prime denote differentiation with respect to $a$ and an Einstein-de
Sitter cosmology has again been assumed. This homogeneous equation has
solutions $x\propto a^n$, where
 \begin{equation}
n=-\fracj14\left(1\pm\sqrt{{16\Phi_0\over(H_0\Delta q)^2}+1}\right).
\end{equation}
 On dimensional grounds $\Phi_0/(H_0\Delta q)^2\sim 1$, so the value of $n$
for the growing solution is 
 \begin{eqnarray}\label{givesn}
n_{\rm g}&=&{\Phi_0^{1/2}\over H_0\Delta q}
-\fracj14+{H_0\Delta q\over32\Phi_0^{1/2}}+\cdots\nonumber\\
&\simeq&{\Phi_0^{1/2}\over H_0\Delta q}-\fracj14.
\end{eqnarray}
 Inserting $x=x_0(a/a_0)^{n_{\rm g}}$, where $x_0$ and $a_0$ are constants
to be determined, into  $p_{\rm r}=a^{3/2}H_0x'$, we find that
the Poincar\'e invariant associated with neighbouring particles is
\begin{equation}\label{givesEtau}
\Gamma=x_0\Phi_0^{1/2}(a/a_0)^{n_{\rm g}+1/2}.
\end{equation}

By Ewald summation we find that the potential (\ref{defsPhi0}) governing the
displacement of a single particle from its position in the lattice has
coefficient
 \begin{equation}
\Phi_0={3\over2\pi}(H_0\Delta q)^2.
\end{equation}
 With this value of $\Phi_0$ equations (\ref{givesn}) and (\ref{givesEtau})
yield 
\begin{equation}\label{unequiest}
\Gamma=\sqrt{{3\over2\pi}}(a/a_0)^{0.94}(H_0\Delta q x_0).
\end{equation}
 If we set $a_0$ equal to the expansion factor at which virialization
occurs, then $x_0$ becomes the magnitude of the particle's displacement at
this epoch, and will therefore be a significant fraction of $\Delta q$.
Consequently, our
new estimate of $\Gamma$ is similar to that of equation (\ref{firstest}).

\subsection{Graininess of the Potential}

Since the gravitational force is calculated from the positions of a finite
number of particles rather than the density field of the true continuum, it
contains spurious high-frequency structure. A lower limit on the
magnitude of this structure is given by the force from the nearest
neighbour -- in a collisionless system, forces from neighbours should always
be small compared to the overall force. The contribution that this makes to
the spurious velocity of a particle is
\begin{equation}
\delta v=\int \d t\,{Gm\Delta x\over a^2[(\Delta x)^2+\epsilon^2]^{3/2}},
\end{equation}
 where we have assumed that Plummer softening is being used and $v$ is the true
peculiar velocity. Substituting for $\Delta x$ from equation (\ref{basice}),
and again assuming a flat cosmology, this becomes
 \begin{eqnarray}\label{grainv}
\delta v&=&{Gm\over H_0(\Delta q)^2}\nonumber\\
&\times&\int_0^{a_0} \!\!\!\!
{\d a\,[1-(a/a_0)\cos(k\Delta q)]
\over a^{1/2}\{[1-(a/a_0)\cos(k\Delta q)]^2+(\epsilon/\Delta q)^2\}^{3/2}}
\end{eqnarray}
 The expression in braces on the bottom of the integrand starts near unity
and tends to $(\epsilon/\Delta q)^2$. Typically the softening length
$\epsilon$ is a significant fraction of $\Delta q$, so the total variation
of this term is not large. The square bracket on the top starts at unity and
vanishes abruptly when $a\simeq a_0$. Hence, the dominant variation of the
integrand comes from the factor $a^{1/2}$ on the bottom, and a reasonable
approximation is
 \begin{equation}
\delta v={2Gma_0^{1/2}\over H_0(\Delta q)^2}.
\end{equation}
 Multiplying by $a_0\Delta x$ we obtain an estimate of the Poincar\'e
invariant generated by graininess of the potential
 \begin{equation}
\Gamma=(\delta v)(a_0\Delta x)
={2Gm\over H_0\Delta q}a_0^{3/2}[1-(a/a_0)\cos(k\Delta q)].
\end{equation}
 In a flat, matter-dominated cosmology, $Gm/\Delta q=(H_0\Delta q)^2/16$,
so this equation can be written
\begin{equation}\label{grainest}
\Gamma=H_0(\Delta q)^2{a_0^{3/2}\over8}[1-(a/a_0)\cos(k\Delta q)],
\end{equation}
 which is very again similar to equation (\ref{firstest}).

\section{Effect of virialization}

\begin{figure}
\psfig{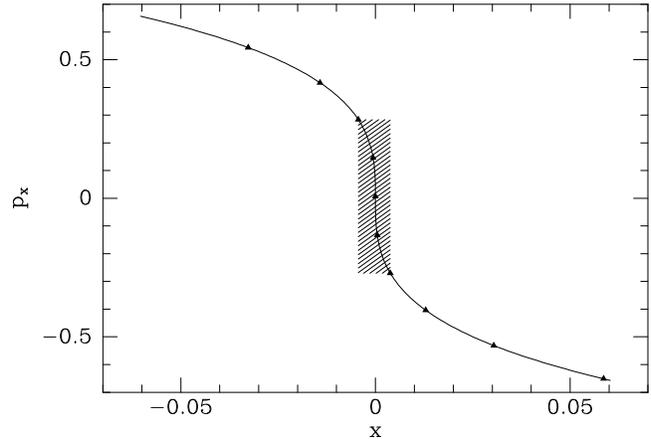}
\caption{Shaded region shows estimate of the Poincare\'e invariant occupied by
the central 5 points after virialization.\label{shadefig}}
\end{figure}

Fig.~\ref{shadefig} shows a phase-space distribution of points that is
rather sparsely sampling a Zel'dovich wave as it collapses. From
this point on the coarse-grained phase-space density will at most points
decline rapidly as the sheet wraps round and round itself, trapping ever
more air in the roll.  So let us ask what the smallest Poincar\'e invariant
is that we could associate with the $n$ particles nearest the centre of the
roll. We take this to be the shaded rectangle. Its invariant is
 \begin{eqnarray}\label{viriest}
\Gamma_n&=&\left(n\Delta q-{\sin(nk\Delta q)\over k}\right)
H_0a^{1/2}_0{\sin(nk\Delta q)\over k}\nonumber\\
&\simeq&H_0(\Delta q)^2\left[{n^4\over3!}a_0^{1/2}(k\Delta q)^2\right].
\end{eqnarray}
 Again the dimensional factor $H_0(\Delta q)^2$ emerges, but this time
multiplied by a numerical factor that is larger by $n^4/3$ than the
corresponding one in equation (\ref{firstest}). This factor is not large for
a sparsely sampled wave, and violent relaxation makes the estimate
inapplicable to the case of a densely sampled wave. We conclude that on
virialization the peak phase-space density does not decline strongly.

\end{document}